\documentclass[%
 reprint,
 amsmath,amssymb,
 aps,
]{revtex4-2}

\usepackage{graphicx}% Include figure files
\usepackage{dcolumn}% Align table columns on decimal point
\usepackage{bm}% bold math
\usepackage{color,soul}
\usepackage{amsmath}
\usepackage[utf8]{inputenc}
\usepackage[english]{babel}
\usepackage{setspace}
\usepackage{color,soul}
\usepackage{xcolor}

\usepackage{chngcntr}
\counterwithout{figure}{section}
\counterwithout{equation}{section}
\begin{document}

%\preprint{APS/123-QED}

\title{High-throughput antibody screening with high-quality factor nanophotonics and bioprinting}

\author{Sajjad Abdollahramezani$^{1\ast}$}
\author{Darrell Omo-Lamai$^{1}$}
\author{Gerlof Bosman$^{2}$}
\author{Omid Hemmatyar$^{1}$}
\author{Sahil Dagli$^{1}$}
\author{Varun Dolia$^{1}$}
\author{Kai Chang$^{3}$}
\author{Nicholas A. Güsken$^{4}$}
\author{Hamish Carr Delgado$^{1}$}
\author{Geert-Jan Boons$^{2}$}
\author{Mark L. Brongersma$^{4}$}
\author{Fareeha Safir$^{5}$}
\author{Butrus T. Khuri-Yakub$^{3}$}
\author{Parivash Moradifar$^{1}$}
\author{Jennifer Dionne$^{1,6}$}

\email{Email: ramezani@stanford.edu, jdionne@stanford.edu}

\affiliation{$^{1}$Department of Materials Science and Engineering, Stanford University, Stanford, CA, USA}
\affiliation{$^{2}$Department of Chemical Biology and Drug Discovery, Utrecht University, Utrecht, Netherlands}
\affiliation{$^{3}$Department of Electrical Engineering, Stanford University, Stanford, CA, USA}
\affiliation{$^{4}$Geballe Laboratory for Advanced Materials, Stanford University, Stanford University, Stanford, CA, USA}
\affiliation{$^{5}$Pumpkinseed Technologies, Palo Alto, CA, Stanford}
\affiliation{$^{6}$Department of Radiology, Stanford University, Stanford, CA, USA}

\date{\today}

\begin{abstract}

Empirical investigation of the quintillion-scale, functionally diverse antibody repertoires that can be generated synthetically or naturally is critical for identifying potential biotherapeutic leads, yet remains burdensome. We present high-throughput nanophotonics- and bioprinter-enabled screening (HT-NaBS), a multiplexed assay for large-scale, sample-efficient, and rapid characterization of antibody libraries. Our platform is built upon independently addressable pixelated nanoantennas exhibiting wavelength-scale mode volumes, high-quality factors (high-Q) exceeding 5000, and pattern densities exceeding one million sensors per square centimeter. Our custom-built acoustic bioprinter enables individual sensor functionalization via the deposition of picoliter droplets from a library of capture antigens at rates up to 25,000 droplets per second. We detect subtle differentiation in the target binding signature through spatially-resolved spectral imaging of hundreds of resonators simultaneously, elucidating antigen-antibody binding kinetic rates, affinity constant, and specificity.  We demonstrate HT-NaBS on a panel of antibodies targeting SARS-CoV-2, Influenza A, and Influenza B antigens, with a sub-picomolar limit of detection within 30 minutes. Furthermore, through epitope binning analysis, we demonstrate the competence and diversity of a library of native antibodies targeting functional epitopes on a priority pathogen (H5N1 bird flu) and on glycosylated therapeutic Cetuximab antibodies against epidermal growth factor receptor. With a roadmap to image tens of thousands of sensors simultaneously, this high-throughput, resource-efficient, and label-free platform can rapidly screen for high-affinity and broad epitope coverage, accelerating biotherapeutic discovery and \textit{de novo} protein design.

\end{abstract}

% \keywords{dynamic metasurfaces, phase-change materials, nanophotonics, structural colors}

\maketitle

% \tableofcontents

% \section*{Introduction}

Antibodies are central components of the adaptive immune system that recognize, bind, and neutralize pathogenic or diseased antigens. The antigen-binding fragment (Fab) of the antibody interacts with the epitope of the exogenous antigen to mediate this critical immune response. Fab comprises a conserved framework interspersed with a hypervariable sequence of $\sim$~60 amino acids, creating a vast humoral repertoire of approximately 10$^{18}$ antibodies \cite{briney2019commonality, carter2018next}. This diversity in sequence and structure underpins antibodies' role as the largest class of biotherapeutics, targeting a broad spectrum of diseases, including infectious diseases, cancer, and autoimmune disorders \cite{pantaleo2022antibodies, paul2024cancer}.

The pursuit of specific, high-affinity, and non-immunogenic biotherapeutics with minimal off-target effects necessitates the generation and characterization of extremely large combinatorial antibody libraries. Current \textit{in vivo} and \textit{in vitro} technologies, including hybridoma technology \cite{kohler1975continuous}, phage display \cite{mccafferty1990phage}, and synthetic libraries \cite{winter2020harnessing}, bolstered by gene editing techniques \cite{notin2024machine, abramson2024accurate}, generate billions of potential therapeutic antibodies \cite{lu2020development}. However, the scale of these libraries exceeds the practical capabilities of existing endpoint binding assays, like enzyme-linked immunosorbent assay (ELISA) \cite{tighe2015elisa}, biolayer interferometry \cite{jug2024biolayer}, and surface plasmon resonance (SPR) methods \cite{homola2003present, mayer2011localized}. To date, these techniques present inherent trade-offs in throughput, sample consumption, and instrumentation/infrastructure. For example, they are generally restricted to the simultaneous analysis of standard well-plate binding interactions (generally 96 or 384 wells) and require several tens to hundreds of microliters of sample per well. Recent advances in automated microfluidics for the quantitative functional characterization of proteins have enabled 1600 simultaneous biochemical measurements \cite{markin2021revealing, delrosso2023large}. However, for the selection of lead antibody candidates, nominally millions of antigen-antibody interactions would be screened. 

Nanophotonic structures enable precise control of the phase, amplitude, polarization, and wavelength of light. In the past few years, they have advanced molecular sensing and spectroscopy by controlling subwavelength light-biomolecule interactions \cite{krasnok2018spectroscopy, tseng2020dielectric, altug2022advances}. These structures transduce molecular interactions occurring within the near-field into free-space optical signals, facilitating multiplexed assays \cite{tittl2018imaging}, single-molecule detection \cite{iwanaga2023metasurface}, and label-free target recognition \cite{hu2023rapid, rosas2023metasurface}. The performance of these sensors is mainly determined by their resonator quality factor (Q, defined as the resonance wavelength divided by the linewidth), mode volume, and controllability over far-field radiation. Each of these factors can be interdependent and impose trade-offs in design optimization. Plasmonic sensors, with their strongly enhanced electromagnetic fields (generally in the range of a few hundreds) and small mode volumes (typically around one-hundredth of the cubic volume of the effective wavelength), mainly suffer from low-Q factors (on the order of a few tens) due to dissipative losses \cite{rodrigo2015mid, oh2021nanophotonic}. In contrast, high-index dielectric nanostructures leverage Mie resonances to minimize nonradiative losses but tend to have reduced field intensities and less effective overlap with target small molecules \cite{yavas2017chip, yavas2019unravelling}. Recently developed  semiconducting metasurfaces supporting quasi-bound states in the continuum (qBIC) modes offer sharp, confined, and surface-sensitive resonances to improve sensing performance \cite{hsu2016bound, koshelev2018asymmetric}. However, qBIC modes generally require extensive two-dimensional nanostructure arrangements with relatively large footprints. This presents challenges such as limited multiplexing, high analyte consumption, reduced sensitivity to small-scale binding events, and increased susceptibility to background noise, defects, and fabrication imperfections \cite{yesilkoy2019ultrasensitive, jahani2021imaging}.

In this work, we develop high-throughput nanophotonics- and bioprinter-enabled antibody screening (HT-NaBS), a multiplexed platform designed for large-scale, low-sample-consumption, and rapid profiling of antibody repertoires. HT-NaBS capitalizes on our lab’s high-Q pixelated nanoantennas \cite{dolia2024very} supporting guided mode resonances (GMRs), with Q-factors exceeding 5000 and densities of 1 million sensors on a 1 cm$^2$ chip. Using digitized acoustic bioprinting, we print capture antigens at rates up to 25,000 droplets per second and achieve site-specific functionalization of each sensor on our chip. Hyperspectral imaging provides rapid optical readouts, allowing the quantification of binding events occurring across hundreds to tens of thousands of sensors simultaneously. We demonstrate the platform's capabilities by measuring the molecular affinity characteristics and selectivity of a panel of neutralizing antibodies targeting SARS-CoV-2, Influenza A, and Influenza B, as well as therapeutic antibodies such as Herceptin and Cetuximab. With a sub-picomolar (sub-pM) limit of detection (LOD), HT-NaBS exhibits a four-order of magnitude linear dynamic range spanning from tens of pM to sub-micromolar (sub-$\mu$M) concentration. We further characterize the binding kinetic rates and affinity constants of each of these antigen/antibody pairs within 30 minutes. Finally, we evaluate the diversity of the antibody epitope landscape, focusing on two classes of antigen-antibody interactions: i) native antibodies targeting distinct functional epitopes of the H5N1 strain of avian Influenza A, and ii) post-translationally modified antibodies targeting epidermal growth factor receptor (EGFR). With the potential for exponentially higher throughput than existing screening assays, HT-NaBS could accelerate antibody-based drug discovery and empirical screening of proteins designed \textit{de novo}.

\section*{Results and Discussions}

HT-NaBS combines silicon nanophotonics with microfluidics and resonant hyperspectral imaging. Prior to antibody screening, acoustic droplet ejection is used for site-specific molecular functionalization of our optical chips. Figure \ref{Fig_1}a illustrates the silicon sensor design, which features a large-scale array of pixelated high-Q nanoantennas. Each nanoantenna is individually functionalized with specific surface chemistries via a custom-built acoustic bioprinter. The automated bioprinter is equipped with a programmable translational stage and enables multiplexed sensing by sequentially depositing different capture antigens on individual nanoantennas. The functionalized chip is integrated with a single-chamber microfluidic cell that has independent inlets for the continuous delivery of target antibodies (see Fig. \ref{Fig_1}b). The optofluidic chip is top-illuminated with a tunable near-infrared (near-IR) light source coupled to a charge-coupled device (CCD) camera to record time-resolved images containing spectral information from hundreds of nanoantennas with a resolution of 10 ms. For real-time binding kinetics, a supercontinuum laser is used for excitation along a collinear optical path, and a spectrometer is employed for the acquisition of reflection spectra from a one-dimensional array of nanoantennas.

\subsection*{High-Q nanoantennas with independent addressability}

\begin{figure*}[htbp]
\centering
\includegraphics[width=1\linewidth, trim={0cm 2cm 9.5cm 0cm},clip]{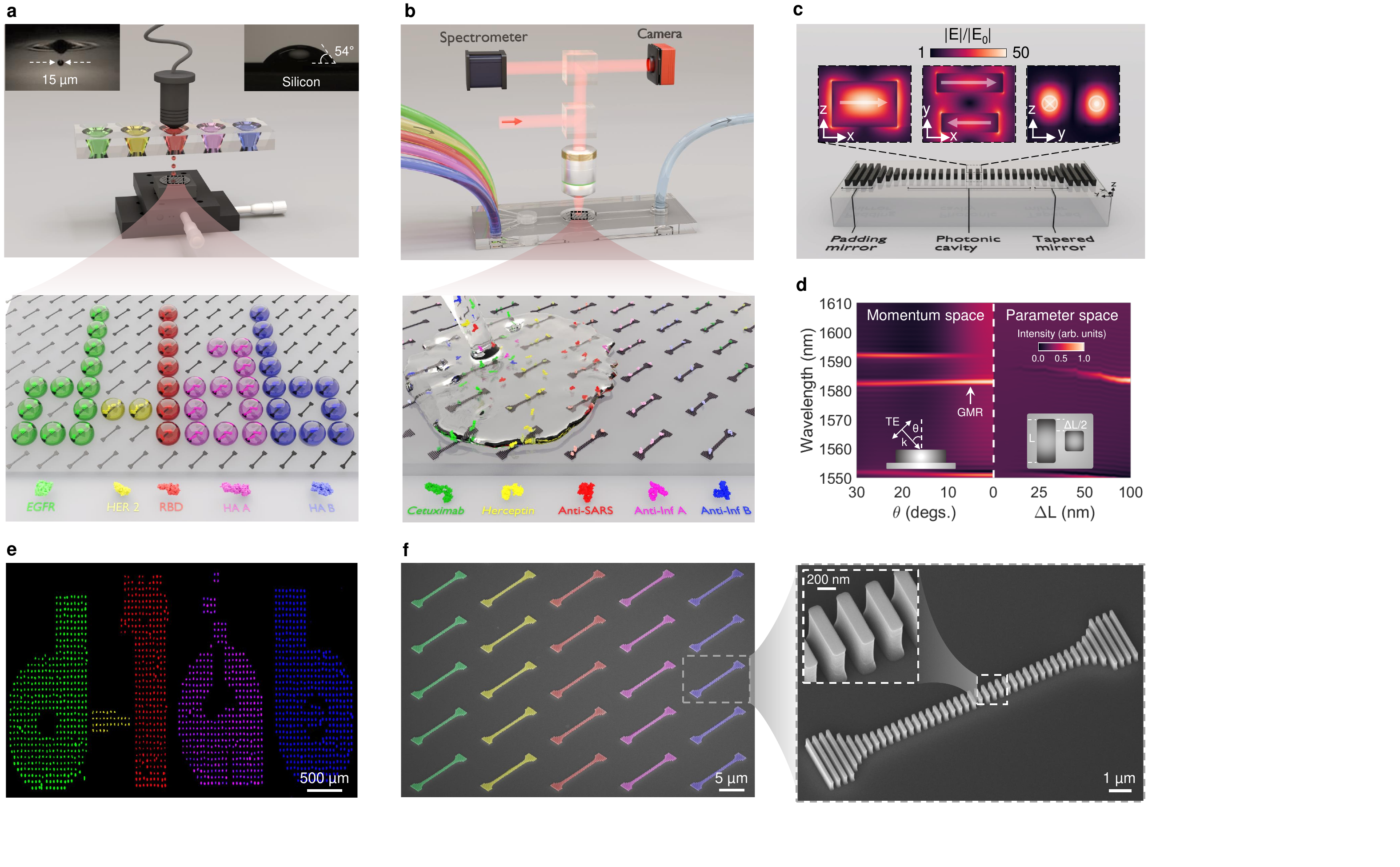}
\caption{\textbf{Overview of HT-NaBS.} 
\textbf{a,} Digitized acoustic bioprinting of biological samples. Schematic of the acoustic bioprinting platform, including an ultrasound transducer, multiwell plates, and a translational stage. Left inset: stroboscopic image of a 15-$\mu$m diameter, $\sim$2-pL volume droplet ejected from a well containing a biostable buffer (see also Supplementary Fig. 6). Right inset: high contact angle between a droplet and the hydrophobic surface of a chip functionalized with epoxysilane SAM (see Supplementary Fig. 12 for comparison with a non-functionalized surface). Zoom-in view: schematic of a two-dimensional array of high-Q nanoantennas, each individually addressed by a droplet ejected from solutions containing color-coded antigens. 
\textbf{b,} Illustration of the optofluidic system integrated with a functionalized chip. Zoom-in view: upon flowing a mixture of antibodies, each color-coded antibody is selectively captured by the relevant immobilized antigen deposited by the bioprinter. 
\textbf{c,} Schematic representation of a silicon nanoantenna. Insets depict the normalized electric field intensity distribution at GMR wavelength and the direction of the induced anti-parallel in-plane dipole moments. The nanoantenna exhibits over 40-fold increase in the electric near-field enhancement and a mode volume close to 1($\lambda$/n$_{\text{eff}}$)$^3$ (see Supplementary Note 1 and Supplementary Fig. 1 for more details). 
\textbf{d,} Numerical analysis of GMR in both momentum and parameter spaces. Simulated reflection spectrum of a nanoantenna with ($\Delta$L) of 100 nm within the cavity section, evaluated in momentum space as a function of incident angle ($\theta$), alongside the reflection spectrum of nanoantennas with varying ($\Delta$L) under normal excitation ($\theta$ = 0$^\circ$) in the parameter space. Despite its high-Q characteristics, our nanoantenna design exhibits minimal sensitivity to changes in the excitation angle. 
\textbf{e,} Tiled fluorescent microscope image of the printed ``d-lab''' displays droplets of EGFR (‘d’), HER2 (‘-’), RBD (‘l’), HA A (‘a’), and HA B (‘b’), as illustrated in \textbf{a}, each conjugated with Alexa Fluor dyes 568 (green), 647 (yellow), 750 (red), 405 (pink), and 488 (blue) dyes, respectively.
\textbf{f,} False-colored tilted SEM image of an array of fabricated nanoantennas and bird’s eye view of an individual nanoantenna with an enlarged SEM of the center of the photonic cavity as the inset. Colors are matched with the target antibodies represented in \textbf{b}.}
\label{Fig_1}
\end{figure*}

Figure \ref{Fig_1}c shows the building block of our nanophotonic sensor: an individually addressable, high-Q nanoantenna, adopted from our “very-large-scale integrated nanoantenna pixel” (VINPix) design \cite{dolia2024very}. This resonant structure is formed from a truncated, one-dimensional, periodic array of symmetry-broken silicon nanoblocks (photonic cavity) supporting GMRs. The periodic modulation of nanoblock lengths, characterized by the perturbation length ($\Delta$L, see inset in Fig. \ref{Fig_1}d), enables wavevector matching between the incident free-space light and the otherwise bound waveguide modes.  The degree of asymmetry, defined by $\Delta$L/L along the cavity, determines the coupling strength of the GMR with free space radiation, and thus the Q-factor. To mitigate the leakage of the confined mode into free space, we include photonic crystal mirrors, composed of both varying- and fixed-size silicon nanoblocks. The varying-size blocks, incorporated on either side of the nanoantenna, shorten the cavity length and reduce the mode volume, enforcing a Gaussian-like field envelope (see Supplementary Fig. 1). This configuration is followed by padding sections at the cavity's ends, consisting of strong mirror elements that enhance mode confinement without degrading the Q-factor. 
% These mirrors function as modulated Bragg reflectors, generating a tunable photonic bandgap within the structure's calculated band diagram \cite{hu2018experimental}. The size of these mirror blocks can be fine-tuned to optimize the intensity of the reflected guided waves back into the cavity. 
The mode profiles, along with a superimposed outline of the silicon nanoblocks shown in Fig. \ref{Fig_1}c, demonstrate significant electric near-field enhancements, exceeding 40x. Unlike conventional Mie-type resonators, which primarily store the electromagnetic field inside the high-index material, our design enhances spatial overlap with the surrounding media, exposing 34\% of the total electric field energy in surface-confined form. The nanoantenna’s ability to localize light into a tightly confined space results in a small effective mode volume, calculated as one cubic wavelength (see Supplementary Note 1).

This particular nanoantenna design facilitates both misalignment-tolerant illumination and wide-field read-out. To evaluate the optical performance of the nanoantenna, we study the (i) out-of-plane symmetry breaking in momentum space by varying the angle of incidence ($\theta$) for a fixed $\Delta$L, and (ii) in-plane symmetry by breaking the structural symmetry under normal incidence (see Fig. \ref{Fig_1}d). Simulation results indicate that for $\Delta$L = 100 nm, the GMR dispersion remains largely insensitive to $\theta$ over a broad range (0$^\circ$ $<$ $\theta$ $<$ 30$^\circ$). However, increasing $\Delta$L under normal illumination leads to a broadening of the linewidth and a blue shift of the GMR, with an associated increase in scattering loss. The mode profile, shown in the inset of Fig. \ref{Fig_1}c, reveals that the induced electric dipole moments are aligned with the polarization of the incident transverse electric (TE) light. The strength of this effective in-plane dipole moment is coupled to the intensity of the projected incident electric field on the nanoantenna surface, which inherently depends on $\theta$. Consequently, as $\theta$ increases, the radiation intensity decreases. On the other hand, as $\Delta$L approaches zero, the radiative decay rate of the GMR under normal illumination approaches zero, resulting in a confined mode with an infinite Q-factor.

\begin{figure*}[htbp]
\centering
\includegraphics[width=0.7\linewidth, trim={0cm 3.7cm 20cm 0cm},clip]{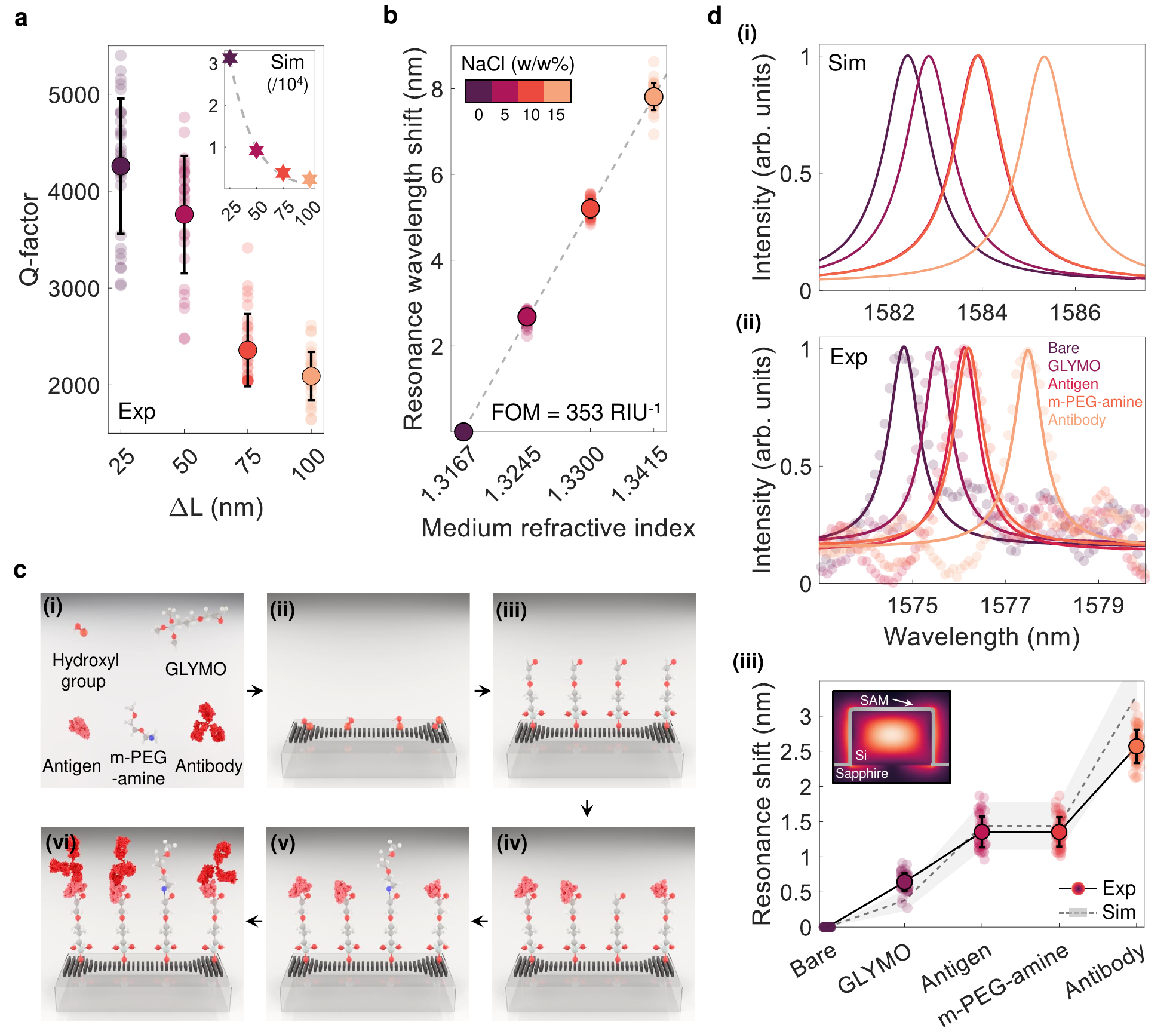}
\caption{\textbf{Design and characterization of high-Q nanoantennas.} 
\textbf{a,} Comparison of experimental (circle markers) and simulated (star markers) Q-factors for different ($\Delta$L) in the cavity section. Mean Q-factor values (bold markers) and standard deviations (error bars) are derived from measurements of 50 nanoantennas for each perturbation condition. 
\textbf{b,} Figure of merit (FOM) of the nanophotonic sensor. GMR wavelength of nanoantenna is measured as a function of the environmental medium's refractive index. The data points represent mean values (bold markers) and standard deviations (error bars), obtained from measurements of 50 nanoantennas at varying concentrations of NaCl. A linear fit to the measured data is used to calculate the FOM. 
\textbf{c,} Multi-step functionalization of nanophotonic sensors. (i) Schematic of chemical components utilized in SAM surface functionalization and proteins (scales are exaggerated here for illustration purposes), (ii) surface hydroxylation, (iii) surface silanization, (iv) immobilization of capture antigens, (v) backfilling with m-PEG-amine to minimize nonspecific binding, (vi) incubation with target antibody. 
\textbf{d,} (i) Simulated and (ii) experimentally measured resonance wavelength shift responses with the sequential addition of each molecular layer represented in \textbf{c}. The resonance wavelength exhibits a consistent redshift corresponding to the cumulative optical mass increase upon the addition of each new molecular layer. (iii) Comparison between the average resonant wavelength shift obtained from simulation (dashed line) and measurements (solid line). Circle markers, bold markers, and error bars represent individual, mean, and standard deviation of measurements from 50 independent nanoantennas. The shaded area represents the resonance shift range, accounting to the minimum and maximum refractive indices of assembled layers in simulations (also see Supplementary Note 2 and Supplementary Fig. 4).}
\label{Fig_2}
\end{figure*}

We validate these simulation results by fabricating an array of pixelated nanoantennas. Figure \ref{Fig_1}f shows a scanning electron microscope (SEM) image of the fabricated chip, along with a bird’s-eye view of an individual nanoantenna. The fabrication process begins by patterning a 600-nm-thick crystalline silicon layer on a sapphire substrate using a standard top-down fabrication method with complementary metal oxide semiconductor (CMOS) compatible techniques (see Methods and Supplementary Fig. 2). The silicon nanophotonic sensor is integrated with a microfluidic chip and characterized under phosphate-buffered saline (PBS) flow. By varying $\Delta$L from 25 nm to 100 nm in four incremental steps, we observe a proportional increase in linewidth, characterized by the full width at half maximum (FWHM), from 0.31 $\mu$m to 0.78 $\mu$m (see Supplementary Fig. 3) and a corresponding decrease in the measured average Q-factor from approximately 4200 to 2100 (see Fig. \ref{Fig_2}a). Theoretically, the Q-factor, dominated by radiative loss, is inversely proportional to the square of the asymmetry parameter (see the inset in Fig. \ref{Fig_2}a). The observed deviations between simulated and experimental results are primarily attributed to radiation losses caused by fabrication imperfections (e.g., surface roughness and statistical disorders) and the non-negligible optical absorption of PBS. With actual dimensions as small as 15 $\mu$m $\times$ 3 $\mu$m, these independently addressable nanoantennas surpass the Q-factors of many extended two-dimensional metasurfaces \cite{liu2019high}. We further evaluate the sensing figure of merit (FOM) of our refractometric platform, defined as sensitivity (resonant wavelength shift per refractive index unit (RIU) change) divided by the GMR’s FWHM. Figure \ref{Fig_2}b shows a FOM of approximately 350 RIU$^{-1}$, achieved by changing the sodium chloride concentration in the background medium to modulate the refractive index. This FOM significantly outperforms plasmonic sensors \cite{yavas2017chip} and even surpasses all-dielectric metasurface sensors \cite{Yang2014all} due to the sensitive high-Q GMR. Combined with the high surface-to-volume ratio of the nanoblocks, which maximizes available binding sites for surface-bound analytes, our design demonstrates significant responsiveness to surface interactions, making it suitable for biorecognition applications.

\subsection*{Self-assembled monolayer functionalization of nanophotonic sensors}

To prepare nanoantennas for antibody detection, we utilize a tailored surface chemistry to selectively bind the target antibody of interest while minimizing nonspecific adsorption of other proteins in the sample. This is achieved through a three-step functionalization process forming a dense and homogeneous self-assembled monolayer (SAM) terminating in a probe, or capture antigen, specific to the target antibody. As illustrated in Fig. \ref{Fig_2}c, the SAM construction begins by treating the nanoantennas with piranha solution to remove organic contaminants and introduce hydrophilic hydroxyl groups (-OH), which are essential for the strong attachment of SAM molecules in the subsequent steps. Next, the nanoantennas are functionalized with 3-glycidoxypropyltrimethoxysilane (GLYMO) (see Methods for more details), where the silane group of GLYMO reacts with the hydroxyl groups on the silicon surface, forming robust silicon-oxygen-silicon bonds, leaving the epoxy group available for further functionalization \cite{tsukruk1999sticky}. After acoustically printing the capture antigens on individual nanoantennas (see below), a covalent nitrogen-carbon bond is formed between the epoxy group and any free amine of the capture antigen at the nanoantenna sites. Given the prevalence of free amines on numerous proteins, this flexible functionalization approach allows for the immobilization of multiple capture antigens across the nanophotonic sensor, facilitating straightforward assay customization and multiplexing. Finally, the chip is incubated with methoxy-PEG-amine (m-PEG-amine), a biologically inert molecule, to block nonspecific binding \cite{obermeier2011multifunctional}. The short PEG molecules, with a molecular weight of 350 Daltons, are small enough to fill the gaps between the relatively large capture antigens (e.g., $\sim$26.5 kDa) and bind to the remaining free epoxy groups, thereby preventing other amine-containing analytes in the multiplexed samples from non-specifically binding \cite{zalipsky1997introduction}. This method provides a robust and adaptable platform for antibody detection, offering high specificity and minimal background interference.

As illustrated in Fig. \ref{Fig_2}d(i), the sequential deposition of GLYMO, antigen, m-PEG-amine, and antibody monolayers results in a consistent redshift in the resonant wavelength across the spectra of individual nanoantennas. The relative shift observed at each layer corresponds to the optical mass of the deposited material. We simulate this molecular assembly by modeling each deposited material as a thin dielectric shell, representing closely packed molecules surrounding the nanoantennas (see Supplementary Note 2 and Supplementary Fig. 4). Fig. \ref{Fig_2}d(iii) demonstrates a clear correlation between the simulated and experimental resonance shifts. The deviation between the experimental and simulated shifts for the GLYMO and capture antigen layer is likely attributable to the tendency of epoxysilane molecules to form non-homogeneous, sparse, and partial SAM \cite{zhang2009one}. The discrepancy between the capture antigen and m-PEG-amine layer is likely due to the possible loss of inadequately bound antigens from the surface caused by insufficient rinsing. Likewise, the variation in antibody binding is likely influenced by steric hindrance, stochastic walking of antibodies, thermodynamic effects, surface repulsion, and the unfavorable orientation of epitopes toward the surface, modulated by the number and spatial distribution of accessible free amines \cite{preiner2014iggs, de2016simulation, hadzhieva2017impact}. These factors can impact the packing density and binding efficiency of antibodies.

\subsection*{Digitized acoustic bioprinting for large-scale, multiplexed registration of nanoantennas}

Spectral responses from the dense array of nanoantennas demonstrate that the performance of high-Q GMR remains unaffected by the proximity of neighboring nanoantennas, down to separations of 3 $\mu$m. This minimal array coupling enables the development of high form-factor nanophotonic sensors with over one million independently addressable sensing elements on a single 1 cm$^2$ chip. To fully exploit the multiplexed sensing capabilities of this platform, we employ acoustic droplet ejection (ADE) technology to selectively deliver biomolecules to individual nanoantennas within the array. 

Acoustic bioprinting is a nozzle-free ADE technique in which a radio frequency signal excites a transducer at its resonant frequency, generating ultrasonic waves that exert force on the free surface of a biofluid \cite{safir2023combining, elrod1989nozzleless}. This force overcomes surface tension, resulting in the ejection of a small droplet. In our assay, the printer reservoir is filled with the fluid of interest, and transducer input parameters are precisely tuned to create localized high-pressure regions within the fluid, leading to controlled droplet formation and ejection. A full description of the electrically driven acoustic-based set-up coupled to a stroboscopic imaging system is provided in Methods and Supplementary Fig. 5. By precisely controlling the electrical pulse parameters (e.g., frequency, amplitude, and width) applied to the transducer, individual droplets as small as 15 $\mu$m (see inset in Fig. \ref{Fig_1}a) can be deposited with deposition rates up to 25 kHz and sub-micrometer precision. As a contactless, nozzle-free technique, ADE enables repeatable deposition of picoliter to nanoliter droplets from low-dead volume (less than 100 $\mu$L) reservoirs without risk of cross-contamination and clogging. The gentle acoustic forces (on the order of tens of N/m$^2$) minimize the potential for damaging delicate biomolecules, preserving their structural integrity and viability \cite{dholakia2020comparing}. 

To validate the high speed and high spatial resolution of the printing process for our molecular assay, we print onto a 10 mm$^2$ array of nanoantennas using multiple capture antigen solutions. As a proof of concept, we illustrate printing of both cancer antigens (EGFR and human epidermal growth factor receptor 2 (HER2)) and infectious disease antigens (SARS-CoV-2, Inf A, Inf B). EGFR and HER2 are transmembrane proteins that regulate cellular proliferation and survival, and are frequently overexpressed or mutated in various malignancies, thus serving as crucial biomarkers in oncology. Additionally, the receptor-binding domain (RBD) of SARS-CoV-2 and the hemagglutinin (HA) protein of Influenza A and B viruses play pivotal roles in mediating viral entry into host cells through interactions with specific surface receptors, making these proteins essential therapeutic targets. First, five different antigens are prepared in PBS solution including EGFR, HER2, RBD, Influenza A HA (HA A), and Influenza B HA (HA B). For visualization in Fig. \ref{Fig_1}e, we tag these antigens with Alexa Fluor 568 (green), 647 (yellow), 750 (red), 405 (pink), and 488 (blue) dyes, respectively. These solutions are sequentially loaded into the bioprinter reservoir, and our color-coded ``d-lab'' logo is printed onto the nanoantenna array (see Fig. \ref{Fig_1}a). A false-colored ``five-tone'' fluorescence image is obtained by overlaying individual mosaic images corresponding to each excitation wavelength. The final image in Fig. 1e (also see Supplementary Fig. 6 and Fig. 7) illustrates the printer's capability to stably and precisely deposit capture antigens.

\subsection*{Rapid and large-scale profiling of antibody libraries}

\begin{figure*}[htbp]
\centering
\includegraphics[width=0.75\linewidth, trim={0cm 7cm 22cm 0cm},clip]{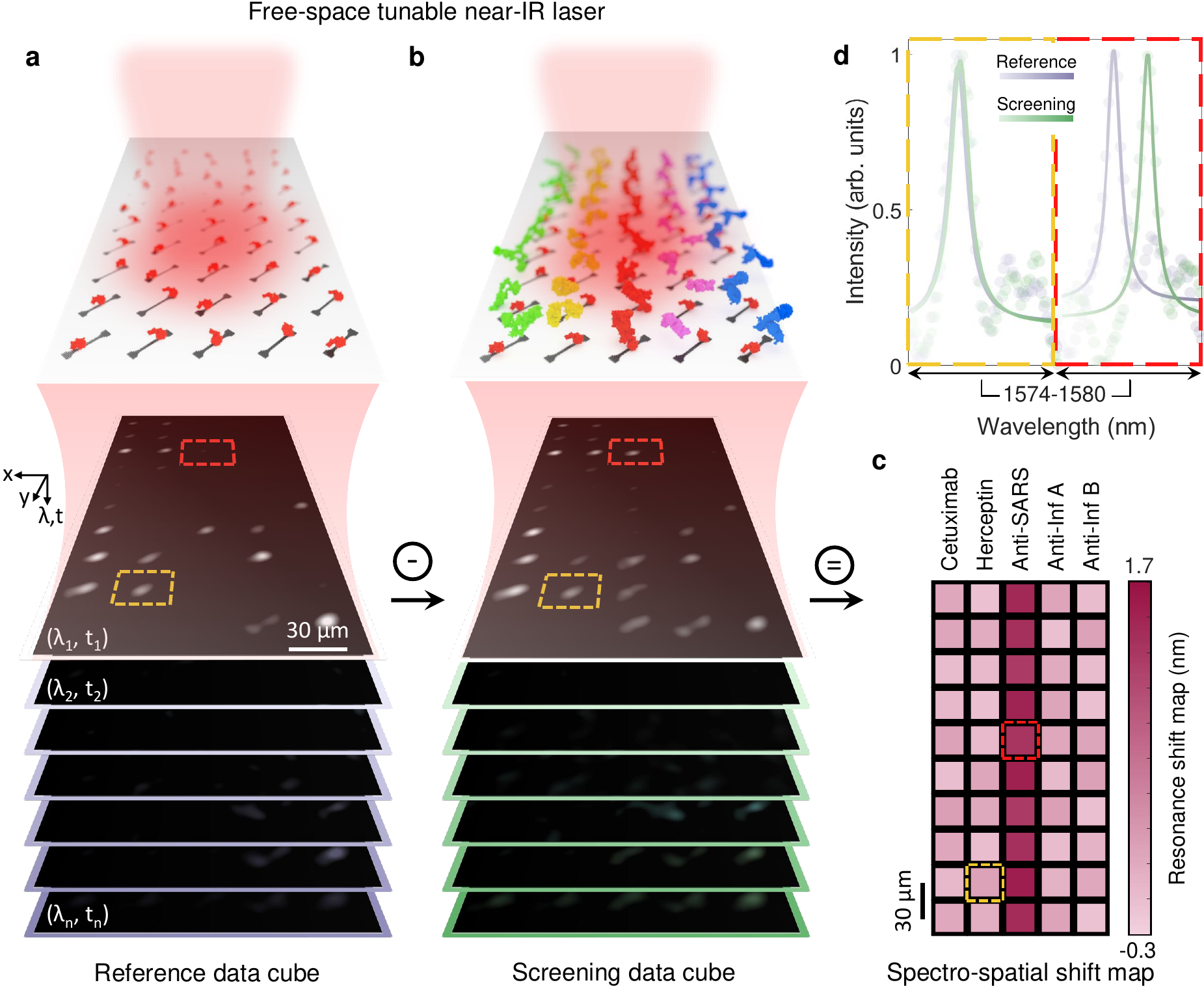}
\caption{\textbf{Working principle of high-throughput antibody screening.}
\textbf{a,b,}. A rendered graphic of the bioassay shows a 10 $\times$ 5 nanoantenna array independently functionalized with RBD (\textbf{a}) before selective deposition of different antibodies with $\muµ$M concentration using digitized acoustic bioprinting (\textbf{b}). Antibodies are color-coded as in Fig. \ref{Fig_1}\textbf{b}. The nanoantenna array is illuminated by a narrow-band tunable near-IR laser, and time-series far-field images from the nanophotonic sensor are captured at each wavelength using a CCD camera. 
\textbf{c,} The spectro-spatial resonance shift map is created by extracting the GMR wavelength of all 50 nanoantennas from the recorded frame of the spectro-microscopy data cube. This is done by subtracting the corresponding image frames in the screening data cube from those in the reference data cube (see Methods for more details). Each block represents processed spatio-spectral information for each nanoantenna. 
\textbf{d,} Extracted spectral data from two nanoantennas, one within (red dashed pixel) and one outside (yellow dashed pixel) the center column, demonstrate the specific binding of anti-SARS-CoV-2 antibodies to the RBD-functionalized surface, while no or negligible binding occurs with non-target antibodies.
}
\label{Fig_3}
\end{figure*}

To measure antigen-antibody interactions, we use hyperspectral imaging to rapidly and simultaneously read hundreds to thousands of individual nanoantennas. We employ wide-field spectro-microscopy (as detailed in Methods) that concurrently resolves both spectral and spatial information enabling multiplexed analyte detection in a single assay, without necessitating a spectrometer \cite{tittl2018imaging}. An inverted reflection microscope, integrated with a tunable, high spectral resolution near-IR laser source, is used to illuminate the optical chip (see Fig. \ref{Fig_1}b and Supplementary Fig. 8). The laser beam is collimated to provide uniform illumination across an array of 50 nanoantennas, though the optical setup could be easily adapted to analyze a significantly larger number (up to $\sim$ 30,000) of nanoantennas (see Supplementary Note 4 and Supplementary Fig. 9). By synchronizing the tunable laser frequency sweep with a 100 frame/s acquisition rate of a two-dimensional CCD array, we acquire a spectro-microscopy data cube, wherein time-resolved frames correspond to the reflectance of the nanoantenna array at specific laser wavelengths (see Fig. \ref{Fig_3}). Using these spectro-spatial maps, we reconstruct the scattering spectra for each nanoantenna by tracking the intensity at a given image pixel, corresponding to the spatial location of individual nanoantennas (see Methods for more details). 

To assess the biosensing capabilities of our platform, we first record the reference ``data cube'' from five columns of nanoantennas, all functionalized with printed RBD of the SARS-CoV-2 spike protein (see Fig. \ref{Fig_3}a). We subsequently acquire the screening ‘data cube’ following the digitized deposition of Cetuximab, Herceptin, anti-SARS-CoV-2 antibody, anti-Influenza A antibody, and anti-Influenza B antibody onto the respective columns (see Fig. \ref{Fig_3}b). Shifts in the spectro-spatial shift map (see Fig. \ref{Fig_3}c) are determined by subtracting the screening maps from the reference maps at each wavelength, spanning the resonances across the entire field of view. As shown in Fig. 3c, an average redshift of approximately 1.5 nm (1.5x the FWHM of the high-Q resonators) in the resonance wavelength is observed in the central column, confirming successful selective binding between the RBD and anti-SARS-CoV-2 antibody. In contrast, no significant shift is detected in other columns, indicating minimal nonspecific binding between the RBD and non-target antibodies. Figure \ref{Fig_3}d illustrates this result by comparing the extracted spectral responses of two nanoantennas, one within and one outside the central column (see dashed boxes in Fig. \ref{Fig_3}b).

\begin{figure*}[htbp]
\centering
\includegraphics[width=0.65\linewidth, trim={0cm 0cm 25cm 0cm},clip]{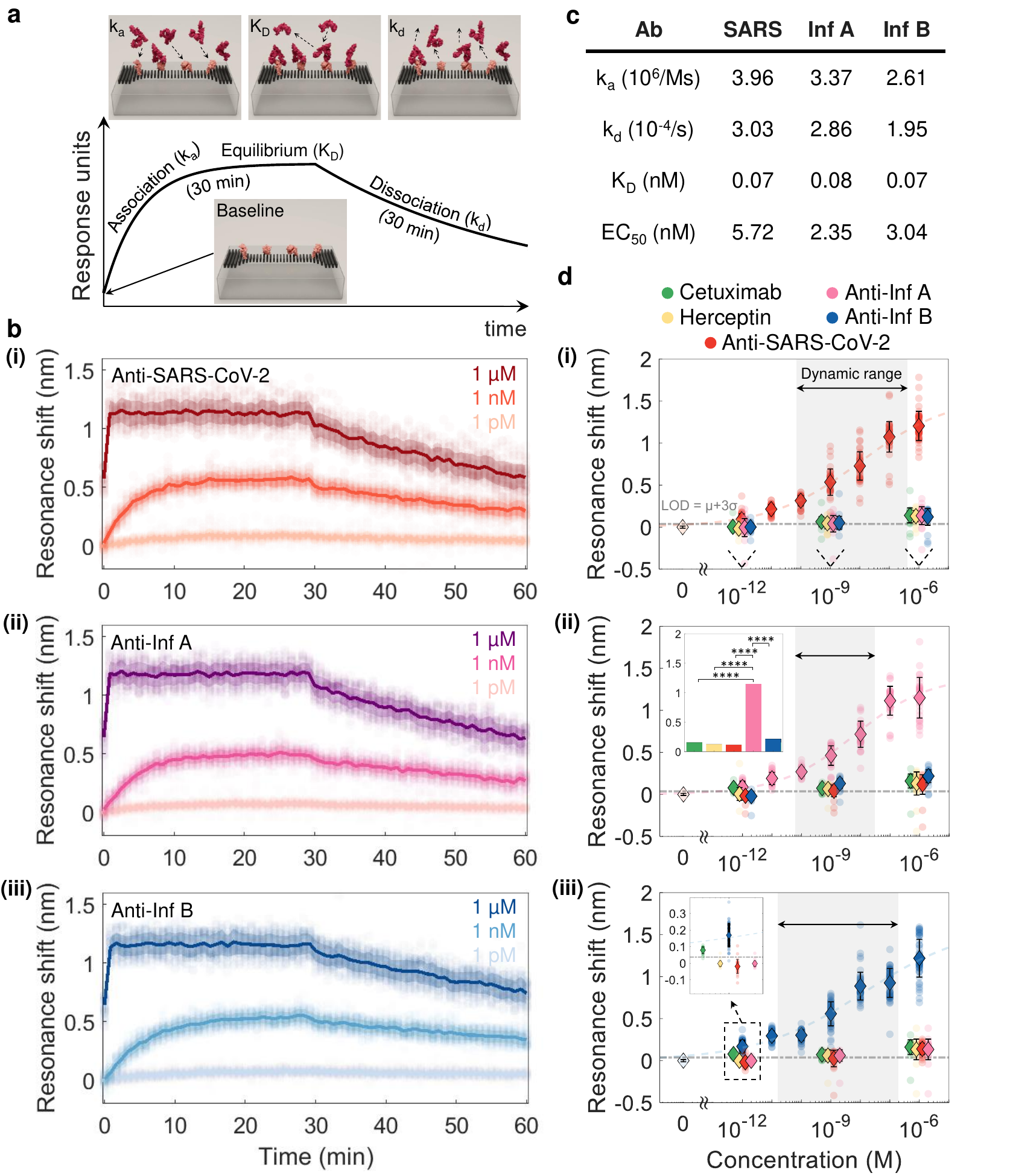}
\caption{\textbf{Binding kinetics, affinity, and specificity measurements of pathogenic antigens.}
\textbf{a,} Ideal sensorgram indicating different phases of antigen-antibody interactions. The binding kinetic measurement begins with the baseline, where antigens are surface immobilized, followed by the association phase, during which target antibodies adsorb onto the antigens, and the dissociation phase where the detachment of antibodies happens. Kinetic parameters are association rate constant (k$_{\text{a}}$), dissociation rate constant (k$_{\text{d}}$), and the calculated affinity constant (K$_{\text{D}}$ = k$_{\text{d}}$ / k$_{\text{a}}$).
\textbf{b,} Measured real-time binding sensorgrams. The binding response, quantified as resonance wavelength shifts, is measured from 30 distinct nanoantennas exposed to serial dilutions (1 $\mu$M, 1 nM, and 1 pM) of antibodies targeting (i) SARS-CoV-2, (ii) Influenza A, and (iii) Influenza B. Each antibody concentration is injected over the corresponding capture antigens, RBD for SARS-CoV-2, HA A for Influenza A, and HA B for Influenza B, deposited on the nanoantennas at a concentration of 1 $\mu$M using the digitized acoustic bioprinter. The markers represent raw data points derived from spectroscopy, while solid lines denote the average response across the replicates. 
\textbf{c,} Binding kinetic values as well as half-maximal effective concentration (EC$_{\text{50}}$) for the affinities between different antibodies and corresponding capture antigens. 
\textbf{d,} Concentration-dependent binding responses are shown for antibodies against SARS-CoV-2, Influenza A, Influenza B, as well as control antibodies Herceptin and Cetuximab, all of which are incubated on nanoantennas functionalized with (i) RBD, (ii) HA A, and (iii) HA B. The minimal resonance wavelength shifts observed for non-targeted antibodies underscore the high specificity of the HT-NaBS platform in high-throughput screening applications for large-scale antibody libraries. Error bars indicate the standard deviations for each target antibody and concentration condition. The LOD is identified by the gray dashed line, representing the mean plus three standard deviations of blank measurements based on IUPAC (International Union of Pure and Applied Chemistry) definition. The color-coded dashed curve corresponds to a fit based on the Hill equation (see Supplementary Note 5). All data points are derived from measurements taken across 30 individual nanoantennas. The dashed area indicates the dynamic range of HT-NaBS. Inset: specificity identification of HA A for anti-Influenza A compared to other antibodies, with ``****'' indicating P $<$ 0.0001 in statistical significance analysis (see Methods for more details). Capture antigen concentrations are 1 $\mu$M.}
\label{Fig_4}
\end{figure*}

Next, we use this setup to analyze binding kinetics and affinities, which are essential characteristics for understanding antibody mechanisms of action and predicting pharmacokinetics and pharmacodynamic profiles \cite{wang2008monoclonal}. The binding strength of an antibody to its antigen is characterized by the association (k$_{\text{a}}$) and dissociation (k$_{\text{d}}$) rate constants, which describe the formation and decay of the antigen-antibody complex. The ratio of these constants, known as the equilibrium affinity constant (K$_{\text{D}}$ = k$_{\text{d}}$ / k$_{\text{a}}$), defines the binding affinity and informs optimal drug dosage recommendations \cite{haes2002nanoscale}. Figure \ref{Fig_4}a illustrates a representative sensorgram of standard affinity biosensor consisting of target antibody association, equilibrium, and dissociation phase.

We quantitatively analyze antigen-antibody interactions for SARS-CoV-2, Influenza A, and Influenza B through real-time, simultaneous optical measurements. As the antibody solution with a specific concentration flows through the microfluidic channel, the GMR peak exhibits a monolithic redshift due to the increased optical mass from the binding effect in proximity to the nanoantennas (see Fig. \ref{Fig_4}b). The sensorgram signal stabilizes once the antigen-antibody binding reaches equilibrium. In the case of Influenza A (Fig. \ref{Fig_4}b(ii)), this condition occurs approximately at the 25, 15, and 1-minute mark following the introduction of antibody solutions with concentrations of pM, nM, and $\mu$M, respectively. At the 30-minute mark, a dissociation cycle is performed using a running buffer to detach bound antibodies from the immobilized antigens (see Methods for more details). As depicted in Fig. \ref{Fig_4}b, the optical response across all target concentrations and across all resonators aligns with standard molecular kinetics and shows excellent concordance with the Langmuir adsorption model (also see Supplementary Note 5) \cite{haes2002nanoscale}. The molecular kinetics characteristics, tabulated in Fig. \ref{Fig_4}c, are consistent with reported values obtained through gold-standard assays, providing further validation of the platform’s accuracy \cite{zost2020potently}. Figure \ref{Fig_4}c also reports EC$_{50}$, representing the concentration of the antibody that produces a response halfway between the baseline and the maximum binding.

To minimize off-target effects, it is essential to prevent non-specific binding of analytes present in complex sample matrices. We assess the specificity of the HT-NaBS platform by investigating the binding interactions between the target and negative control antibodies and a specific capture antigen. In our initial experiment, the RBD is first immobilized on 50 individual nanoantennas via a bioprinting process. We incubate our chip with the antifouling agent, m-PEG-amine, to prevent nonspecific adsorption or binding (see Supplementary Note 3 for details). We then perform a series of optical measurements following sequential incubation of the sample with different concentrations of a single antibody. The study commences with non-target antibodies (Cetuximab, Herceptin, anti-Influenza A, and anti-Influenza B) before testing the target antibody, anti-SARS-CoV-2. As shown in Fig. \ref{Fig_4}d(i), the experimental data reveal a significant spectral shift in the resonance for the target antibody, while negligible shifts are observed for non-target antibodies across a concentration range from pM to $\mu$M. The concentration dependent resonant wavelength shift curve fits the Hill equation, describing target antibody binding coverage (see Supplementary Note 5). To further validate the platform's specificity, similar experiments are conducted with two highly pathogenic antigens, HA A and HA B. As Figs. \ref{Fig_4}d(ii,iii) indicate the results consistently demonstrate the high specificity of HT-NaBS, regardless of the protein of interest. The specificity analysis result presented in the inset of Fig. \ref{Fig_4}d(ii) exhibits high statistical significance (****, p $<$ 0.0001 (see Methods for more details)), indicating that the binding of the anti-Influenza A antibody to HA A is significantly stronger compared to non-target antibodies. This result underscores the specificity of HT-NaBS, as reflected in the substantially higher binding signal.

In addition to specificity, a low LOD and a wide dynamic range are critical for the accurate identification and characterization of lead antibodies with optimal binding affinities and functional properties from a large candidate pool. A low LOD enables the detection of trace amounts of target antibodies, facilitating the evaluation of antibody efficacy at minimal concentrations. Meanwhile, a wide dynamic range ensures that the platform can reliably quantify both low and high antibody concentrations, which is crucial for elucidating dose-response relationships. As illustrated in Fig. \ref{Fig_4}d, HT-NaBS exhibits a linear, near four orders of magnitude dynamic range from (70 pM to 400 nM in Fig. \ref{Fig_4}d(iii)) with an approximate LOD of 45 fM, which is comparable to that of gold-standard affinity-based assays such as ELISA \cite{liu2021covid}, though HT-NaBS offers considerably reduced sample reagents and higher throughput. We note that the LOD could be significantly improved (e.g., down to the aM level), with the high-Q of our resonators and with modified surface chemistry. The observed standard deviation in the resonant wavelength shifts at different concentrations is likely attributable to variability in the orientation of capture antigens on the surface of the nanophotonic sensor, as previously discussed. Moreover, potential blue shifts, possibly caused by the loss of weakly bound or physiosorbed antigens, may also contribute to the observed deviations. These sources of noise can be mitigated by optimizing surface functionalization stability and homogeneity, improving rinse protocols, and selecting more oriented surface functionalization strategies.

\subsection*{High-throughput epitope binning of antibodies against H5N1 avian Influenza and EGFR}

\begin{figure*}[htbp]
\centering
\includegraphics[width=1\linewidth, trim={0cm 1.5cm 8.5cm 0cm},clip]{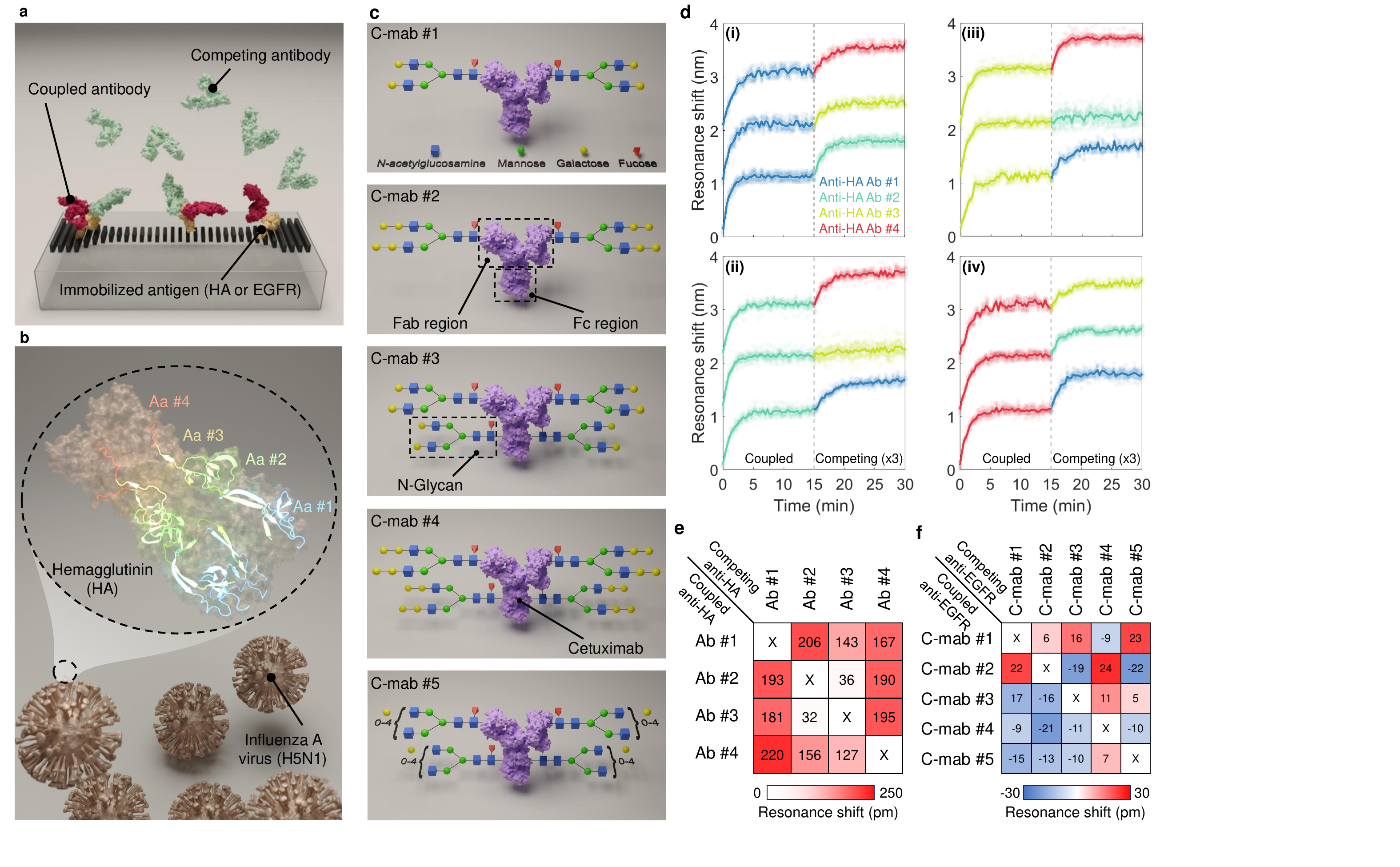}
\caption{\textbf{Epitope binning analysis of antibodies using HT-NaBS.}
\textbf{a,} Cartoon illustration of the tandem assay format. This competitive assay involves the sequential application of competing antibody variants over an immobilized antigen that has been pre-bound with a coupled antibody. The HA (for \textbf{b}, \textbf{d}, and \textbf{e}) or EGFR (for \textbf{c} and \textbf{f}) antigen is deposited onto nanoantennas using the bioprinter. The binding competition between the coupled and competing antibodies (anti-HA Ab or Cetuximab) reveals overlapping or distinct epitopes based on the observed binding interactions. 
\textbf{b,} Schematic representation of the H5N1 strain of avian Influenza A virion alongside the crystal structure of its surface protein HA. Residues are colored according to amino acid diversity; Aa \#1 (amino acids 110-160), Aa \#2 (amino acids 250-310), Aa \#3 (amino acids 260-320), and Aa \#4 (amino acids 320-370). Overlap between the green and yellow sequences resulted in a bright light green color. Control antibodies anti-HA Ab \#1 to Ab \#4 are generated to target relevant immunogenic regions within these specified amino acid sequences. 
\textbf{c,} The three-dimensional structure of a glycoengineered Cetuximab molecule (C-mab) illustrates site-specific glycosylation at both the Fab and Fc regions. Variants of glycoengineered C-mabs are depicted in order of increasing N-glycan complexity, by adding or removing monosaccharides from top to bottom.
\textbf{d,} Competitive binding sensorgrams. A resonance redshift indicates the binding of a competing antibody to an unoccupied epitope on the HA antigen, signifying the presence of non-overlapping epitopes. Conversely, the absence of binding (negligible resonance shift) suggests that the competing antibody’s epitope overlaps with that of the coupled antibody, resulting in competitive blocking. For instance, in panel (ii), anti-HA Ab \#1 and Ab \#4 are shown to bind distinct epitopes from the coupled antibody Ab \#2, whereas Ab \#3 shares an overlapping epitope with Ab \#2. Curves are vertically offset by 1 nm for clarity. To enhance clarity, the resonance shift values for the competing antibodies are scaled by a factor of 3. 
\textbf{e,} Heat map visualization of the epitope binning results for anti-HA variants, where a 4 $\times$ 4 matrix illustrates the pairwise blocking patterns. The color-coded values indicate the resonance shift between each pair of competing and coupled antibodies. White cells denote antibodies that cross-block each other (overlapping epitopes), while reddish cells indicate orthogonal epitope coverage (non-overlapping epitopes).
\textbf{f,} Heat map visualization of anti-EGFR cetuximab variants. Small resonance shifts, derived from measurements of 40 nanoantennas, suggest that all glycoengineered antibodies share identical or closely overlapping epitopes. Resonance shifts for all antibody pairs are provided in Supplementary Fig. 13.}
\label{Fig_5}
\end{figure*}

Determining the binding site of antigens, or epitope, targeted by an antibody is crucial for comprehending its therapeutic mechanism of action and optimizing the discovery of biotherapeutics \cite{harvey2021sars, barnes2020sars}. Unlike affinity, the antibody epitope is an inherent property that cannot be rationally enhanced through engineering methods and ultimately depends on empirical selection \cite{cho2021bispecific, jones2021neutralizing}. Characterizing and grouping a library of antibodies by their binding regions against a specific antigen is essential to mitigate the risk of favoring redundant antibody clones and losing epitope diversity, which can occur when relying solely on affinity selections. High-throughput organization of antibodies into epitope families or ``bins'' facilitates the preservation of epitope diversity and provides critical information for expanding intellectual property protection. In the process of epitope binning, antibodies are tested in a pairwise combinatorial manner (see Fig. \ref{Fig_5}a), and those that compete for the same or closely overlapping binding region are grouped into bins, thereby ensuring a comprehensive understanding of their binding characteristics.

The H5N1 strain of avian influenza A, responsible for a recent dairy cattle outbreak in the U.S. \cite{caserta2024spillover}, is on the World Health Organization’s list of priority pathogens due to its high transmissibility, virulence, and limited access to vaccines and treatments, posing a potential global health threat \cite{mallapaty2024pathogens}. Accordingly, we select a panel of four antibodies (anti-HA Ab \#), each raised against a synthetic peptide corresponding to specific regions of the HA protein of the highly pathogenic H5N1 subtype. These peptides are 10 to 14 amino acids in length and correspond to the amino acid sequences 110-160, 250-300, 260-310, and 320-370, respectively (see Fig. \ref{Fig_5}b). In this study, we employ a tandem epitope binning assay wherein the capture antigen H5N1 HA is first deposited by the acoustic bioprinter and covalently coupled to the nanoantennas (see Fig. \ref{Fig_5}a). After backfilling with m-PEG-amine to minimize nonspecific adsorption, the first antibody (coupled antibody) is injected under saturating antigen-binding conditions using a microfluidic cell, and the binding kinetic response is measured against the antigen alone. Subsequently, washing with the pulse of the running buffer is performed followed by the injection of the second antibody (competing antibody) to quantitatively assess its binding to the existing antigen-antibody complex. The competing antibody's blocking indicates shared binding epitopes, while its ability to bind to the antigen reveals non-overlapping binding sites. Each binning experiment cycle involves a 15-minute injection of the first antibody at a 20-nM concentration, followed by a 15-minute injection of the competing antibody.

The processed molecular kinetics sensorgrams for the antibody panel are shown in Fig. \ref{Fig_5}d. Figure \ref{Fig_5}d(ii) shows antibodies 1 and 3 share the overlapped epitope, as a negligible resonance shift is detected after introducing the latter on the former. However, the pairwise experiments reveal the rest of the panel binds to different epitopes, with over hundred picometer shifts upon introduction of the competing antibody. We summarize the results in the form of a heat map (see Fig. \ref{Fig_5}e). Each color-coded box in this visualization represents the strength of binding between the competing antibody flowing as an analyte (listed horizontally) on the coupled antibody (listed vertically). A white/reddish box thus represents non-/mutual overlapping epitopes between the corresponding antibodies listed in the heat map. The integration of kinetics with epitope binning provides a stronger path for selecting a diverse set of high-affinity antibodies for the target antigen in a single experiment with minimal biosample consumption.

As a second example, we consider EGFR, a key protein in cell growth and proliferation, with its overexpression or mutation contributing to the progression of various cancers \cite{graham2004cetuximab}. Glycoengineering, the process of modifying glycosylation patterns at both the Fab arm and the crystallized fragment (Fc) stem of therapeutic antibodies (see Fig. \ref{Fig_5}c), holds promise for enhancing the efficacy of cancer therapy. Cetuximab, as the only approved anti-EGFR antibody with Fab glycosylation \cite{saporiti2024silico}, serves as an ideal model for investigating these modifications. To elucidate the role of Fab glycosylation in antigen-antibody interactions, we perform epitope binning on a panel of Cetuximab variants with homogeneous glycoforms. Our aim is to characterize the impact of different Fab glycoforms on antigen recognition, specifically identifying potential epitope masking or modulation due to glycan-induced conformational changes.

We employ a panel of five glycoengineered Cetuximab variants with progressively complex glycosylation profiles (see Fig. \ref{Fig_5}c). C-mab \#1, representing a base Fab-glycosylated Cetuximab, followed by C-mab \#2, which introduces an additional glycosylation site on the Fab region to evaluate the effect of increased Fab glycosylation density. C-mab \#3 and C-mab \#4 incorporate Fc glycosylation, while maintaining the Fab glycosylation of C-mab \#1 and C-mab \#2, enabling analysis of the interaction between Fab and Fc glycosylation. Finally, C-mab \#5 represents the most extensively glycosylated variant.

Using pairwise comparisons of these variants, we construct a heatmap blocking map (Fig. \ref{Fig_5}f) based on resonance shifts observed upon the introduction of competing C-mabs. The negligible resonance shift of all boxes indicates that our glycoengineered Cetuximab variants share a common epitope, clustering into a non-overlapping bin. These findings demonstrate that Fab and Fc glycosylation modifications do not significantly affect antigen binding, with all variants targeting the same epitope. Therefore, Fab glycoengineering strategies can be pursued to enhance therapeutic properties, spanning improved aggregation resistance, solubility, thermostability, in vivo half-life, and drug payload conjugation \cite{giddens2018site, jefferis2009glycosylation, van2018adaptive}, without compromising immunoregulatory functions.  

\section*{Conclusion}

HT-NaBS offers a high-throughput, CMOS-compatible solution for antigen-antibody profiling. Our optofluidic chip features high-Q nanoantennas patterned at densities of 1 million per cm$^2$, each with Q-factors of $\sim$ 5,000. Through digitized acoustic bioprinting, we enable precise functionalization of each resonator, facilitating simultaneous quantification of binding events across hundreds to thousands of nanoantennas using hyperspectral imaging. We demonstrated HT-NaBS by profiling antibodies against SARS-CoV-2, Influenza A \& B, and therapeutic targets such as EGFR and HER2, achieving picomolar sensitivity with a near four-order of magnitude linear dynamic range. We also demonstrated the precise characterization of binding kinetics through real-time opto-fluidic measurements, including the determination of k$_{\text{a}}$, k$_{\text{d}}$, and K$_{\text{D}}$. Finally, we demonstrated the utility of HT-NaBS in differentiating shared and unique epitopes, including antibodies against distinct H5N1 epitopes and glycoengineered Cetuximab antibodies against EGFR.  

We note several improvements in HT-NaBS that could enable even higher-throughput and lower reagent-use screening. Foremost, the integration of higher-density pixel imaging platforms could enable simultaneous imaging of hundreds of thousands of resonators. Additionally, acoustic droplet ejection allows the use of multiple printing heads to simultaneously address different nanoantennas with varied fluid samples, enabling increased multiplexing against up to thousands of bioprinted targets. Third, advances in multi-channel microfluidics can achieve numerous parallel microreactors and droplet generators on a single chip, allowing for the independent and simultaneous delivery of target analytes. Finally, to further reduce reagent consumption, designs of pixelated nanoantennas with Q-factors in the hundreds of thousands to millions (e.g., through alternative symmetry breaking mechanisms) would be necessary. Such Q-factors could enable near-single-molecule sensitivity akin to integrated high-Q microcavities and photonic crystals, yet benefiting from free-space excitation and readout. 

With such engineering advances, HT-NaBS stands to offer significantly higher throughput within a reduced footprint, comparable to executing over 10,000 96-well plate assays within a 1 cm$^2$ area. Importantly, various probe molecules (e.g., nucleic acids, aptamers, metabolites, and proteins) can be functionalized onto our surfaces using nozzle-free bioprinting. Therefore, HT-NaBS could serve as a universal photonic microarray for targeting diverse biomolecules. This silicon biophotonic chip coincides with recent breakthroughs in computational protein design, which have facilitated the \textit{in silico} generation of vast libraries of synthetic antibodies and proteins. However, the efficient screening of these extensive datasets to identify candidates with optimal binding properties remains a major challenge. Our findings suggest that HT-NaBS is well-suited to address this bottleneck, thereby accelerating the iterative, lab-in-the-loop process of generative protein design and bridging the gap between computational designs and their translation into therapeutic applications. By generating high-content information on antibody affinity, HT-NaBS may ultimately drive AI-enabled innovation in antibody discovery and biotherapeutic development.

\section*{Methods}

\subsection*{Numerical simulation}

Electromagnetic wave simulations are performed using the Lumerical finite difference time domain (FDTD) solver and COMSOL Multiphysics based on finite element method. Silicon (n = 3.48) nanoantennas are simulated on a sapphire (n = 1.75) substrate with water (n = 1.33) as the environment. Isotropic material properties are used in all cases. Perfectly matched layer boundaries are applied in the x, y, and z directions. Anti-symmetric and symmetric boundary conditions are used in the x and y directions respectively (see Supplementary Fig. 10). A TE-polarized plane wave source injected in the +z direction through the substrate is used to excite the nanoantenna. A rectangular mesh with a maximum dimension of 10 nm in the x, y, and z directions is applied to the structures. The quality factors of the simulated resonance modes are calculated from time domain field decay rates using Lumerical’s Q analysis object, wherein the quality factor is evaluated as Q = $-2\pi f_{R}$log$_{10}(e)/m$, where f$_{\text{R}}$ is the resonant frequency of the mode and m is the temporal field decay rate. To estimate the resonant wavelength shifts associated with the successive molecular layers deposited on the nanoantenna, each surface step is modeled as a thin dielectric shell, and the response is numerically calculated using FDTD simulations (see Supplementary Note 2). 

\subsection*{Device fabrication and material characterization}

The nanoantenna arrays are fabricated using standard lithographic processes (see Supplementary Fig. 2). First, 600 nm single-crystal silicon-on-sapphire substrates (University Wafer) are cleaned by sonication in and rinsing with acetone, methanol, and isopropanol and baked at 180 °C for 2 min. Subsequently, a thin film of hydrogen silsesquioxane (HSQ) negative tone resist (XR-1541-06, Corning) is applied to the substrates via spin-coating at 1000 rpm and baked at 80 $^\circ$C for 5 min. The substrate is then spin-coated with a charge dissipation layer (e-spacer, Showa Denko) at 2000 rpm and further baked at 80 °C for 5 min. The HSQ resist is patterned using an electron beam lithography system (Raith EBPG 5200+) with a 100 kV accelerating voltage and developed in an aqueous solution of 1 wt\% sodium hydroxide and 4 wt\% sodium chloride. The resist pattern is transferred to the silicon substrate by inductively coupled plasma reactive ion etching (Oxford III-V Etcher) with hydrogen bromide and chlorine gasses. The silicon nanoantennas are patterned at 45° to the c-axis of the sapphire substrate to minimize the polarization dependence of their wavelength response. Lastly, the exposed resist is stripped in a 50:1 hydrofluoric acid solution, and the chip is cleaned in Piranha solution (9:1 sulfuric acid:hydrogen peroxide) heated to 120 °C to remove organic residues. Sample images are obtained using an FEI Magellan 400 XHR scanning electron microscope equipped with a field emission gun source. To mitigate charging effects, a $\sim$~7-nm gold-palladium coating is sputtered on the sample surface. Images are acquired under an accelerating voltage in the range of 5-10 kV.

\subsection*{Optical characterization}

Optical measurements of the nanoantennas are performed using a custom-built near-IR reflection setup (see Supplementary Fig. 8). The optical chip is illuminated through the sapphire substrate by a broadband supercontinuum laser (NKT SuperK EXTREME) delivering near-IR broadband light (1180-2400 nm) through a collimated fiber output connected to a spectral splitter (NKT SuperK Split). A cross-polarization strategy is adopted to suppress Fabry–Pérot resonances and reduce background noise. A linear polarizer, s-pol, is positioned to polarize the incident beam at 45° to the nanoantennas. The polarized beam is focused onto the back focal plane of a 5x objective lens (Mitutoyo Plan Apochromat near-IR, NA = 0.14, f = 200) using a bi-convex lens (f = 200 mm) such that a collimated quasi-plane wave is incident on the optical chip. The scattered light collected by the objective lens is passed through a linear polarizer, p-pol, with a -45$^\circ$ polarization angle to the nanoantennas and focused through a bi-convex lens (f = 75 mm) onto the entrance slit of a spectrometer (Princeton Instruments SPR-2300) with a 600 gr/mm diffraction grating. The signal is then focused onto a TE-cooled InGaAs CCD detector (NiRvana, Princeton Instruments). For spectro-microscopy, the optical chip is illuminated via a tunable single frequency laser source (Santec TSL-550-C) with a collimated fiber output (see Supplementary Fig. 8). Optical measurements are performed by sweeping the tunable laser frequency and synchronizing the CCD detector acquisition rate at 100 frame/s. A time series of wide-field (320 $\mu$m $\times$ 160 $\mu$m) image frames are captured where each frame represents the intensity map of a cluster of individual nanoantennas at a specific illumination wavelength. This series forms the spectro-microscopy data cube from which the resonance shift map is calculated.

\subsection*{Resonance fitting and Q-factor extraction}

In this study, the normalized experimentally measured reflectance are analyzed by fitting the resonant spectral feature using

\vspace{0.2cm}
$R = \left| \frac{1}{1 + F \sin^2(n_s k h_s)} \right| \left| a_r + a_i i + \frac{b}{f - f_0 + i \gamma} \right|^2$.
\vspace{0.2cm}

The first term describes the Fabry–Pérot interference through the substrate, characterized by its thickness h$_{\text{s}}$ and refractive index n$_{\text{s}}$, where k represents the free-space wavevector ($2\pi/\lambda$) and F accounts for the reflectivity at the interfaces. The second multiplicative term models the superposition of a constant complex background, a$_{\text{r}}$ + a$_{\text{i}}$ i, with a Lorentzian resonance defined by a resonant frequency f$_0$ and a FWHM of ($2\gamma$). $\gamma$ denotes the overall damping rate of the resonance. The Q-factor is given by Q = f$_0/2\gamma$.

\subsection*{Data acquisition and resonance map computation}

We use MATLAB, Mathworks, R2022a to pre-processing begins by correcting each frame in the recorded hyperspectral data cube for spatial and spectral inconsistencies introduced by the light source. The intensity of each pixel on the CCD camera is normalized by dividing the pixel values from frames collected in regions with and without the nanoantenna array on the substrate, ensuring that identical acquisition parameters are maintained. To capture spatial information, the intensity frames are aggregated into a single composite frame, wherein the local intensity maximas correspond to the centers of the labeled nanoantennas. Since the far-field scattering from each nanoantenna is distributed across a 9 $\times$ 9 pixel grid on the CCD, the spectral data are derived by integrating the intensities of the corresponding 81 pixels. Resonance shift maps are computed across different time points by subtracting the spectro-spatial resonance maps (e.g., post-antibody binding) from reference maps (e.g., post-antigen binding) at the same location on the optical chip.

\subsection*{Statistical analysis}
One-way ANOVA is conducted in MATLAB, Mathworks, R2022a to assess statistical significance among antibody groups, with p-values $<$ 0.05 considered significant. When applicable, pairwise comparisons using Tukey–Kramer HSD are performed to identify significantly different group means.

\subsection*{Sample preparation and surface functionalization}
To achieve SAMs with minimal nonspecific binding, the silicon surface is initially activated for functionalization by immersing the chips in a Piranha solution (9:1 H$_2$SO$_4$:H$_2$O$_2$) heated to 120 $^\circ$C for 60 minutes to hydroxylate the surfaces. Subsequently, the surface is silanized by immersing the chips in a glass vial containing 20 mL of a 1.2\% v/v solution of GLYMO ((3-glycidoxypropyl) trimethoxysilane (3-GPS)) (Sigma Aldrich, 440167) in anhydrous toluene (Sigma Aldrich, 244511). The vial is backfilled with argon, sealed with parafilm, wrapped in aluminum foil to prevent light exposure, and left at room temperature for 24 hours. Following that, the samples are soaked in fresh anhydrous toluene for 5 minutes, sonicated in anhydrous ethanol (Sigma Aldrich, 459836) for 5 minutes, soaked in fresh anhydrous ethanol for 3 minutes, and dried with argon gas. The samples are then heated in a furnace at 110 $^\circ$C for 30 minutes allowing for the desorption of poorly bound molecules and leaving a epoxy-functionalized silane layer. Functionalized samples are subsequently stored in a vacuum environment and should be utilized within a two-week timeframe. For high concentration sensing experiments, optical chips are reused and Piranha cleaned between experiments before repeating the silanization procedure. Fresh chips are used for most low-concentration measurements to avoid recycling effects.

\subsection*{Digitized acoustic bioprinting}
Acoustic bioprinting is conducted using a custom-built ultrasonic immersion transducer (see Supplementary Fig. 5) that operates at a center frequency of 147 MHz and a focal distance of 3.5 mm. The droplet generation process is driven by a waveform generator (Keysight 33600A Series Trueform), which produces a square-wave burst at a continuous repetition frequency of 1 kHz, with a pulse width of 5.5 $\mu$s and a voltage of 1.5 V, sufficient to activate the RF synthesizer (Fluke 6062A). The synthesizer generates a sinusoidal wave at 147 MHz and the desired voltage, which is then amplified by a power amplifier (Minicircuits ZHL-03-5WF+) before reaching the transducer. These parameters are determined using a network analyzer (Hewlett Packard 8751A) and pulse-echo measurements obtained from an oscilloscope (Keysight InfiniiVision DSOX3054A). The transducer is attached to a quartz, spherical focusing lens and mounted on a manual translation stage, positioned above a 303 stainless steel reservoir with a 1 mm aperture.  Solution of interest is pipetted into the gap between the focusing lens and the ejection plate, held in place by surface tension. Stable droplets are ejected with an energy of 0.096 $\mu$J onto the selected substrates. The ejection process is monitored with a high-resolution camera (Allied Vision Guppy Pro F-125 CCD Monochrome Camera) equipped with a 20x objective, focused on the underside of the ejection plate. The camera is positioned opposite a strobing LED, synchronized with the waveform generator. Additionally, the acoustic echo is continuously monitored using an inline oscilloscope (Keysight InfiniiVision DSOX3054A). Aligned printing is performed using an acrylic slide equipped with a 0.5 cm machined hole, mounted onto a custom 3D-printed substrate holder. This holder is attached to two perpendicularly stacked 100 mm brushless DC linear translation stages (Thorlabs DDS100M), controlled by two K-Cube brushless DC servo drivers (Thorlabs KBD101). The optical chip is positioned on the acrylic slide, centered over the hole to allow image capture from below. The substrate is positioned approximately 1 mm below the ejection plate to minimize droplet displacement before reaching the substrate. For alignment, a color camera (Daheng Imaging) and a 20x objective are mounted directly beneath the substrate and focused on the nanoantennas. For precise deposition of individual droplets onto individual nanoantennas, the acoustic printer's focal point is first aligned directly above the center of a patterned marker as a reference. A MATLAB script is linked with Thorlabs APT User Utility software to deposit droplets onto the substrate by simultaneously controlling the motorized stages and the waveform generator. This setup enabled precise triggering of droplet ejection at designated nanoantennas locations.

\subsection*{Preparation, immobilization, and flow dynamics of protein solutions}

To prepare capture antigens, including RBD (AcroBiosystems, SPD-C52H3), Influenza A HA (AcroBiosystems, HA2-V52H7), Influenza B HA (AcroBiosystems, HAE-V52H4), EGFR (AcroBiosystems, EGR-H5222), and HER2 (AcroBiosystems, HE2-H5225), ultrapure DNase/RNase-free distilled water (Thermo Fisher, 10977015) is used for reconstruction. Antigen solutions at various concentrations are prepared in PBS 1x buffer (Thermo Fisher, 10010049) supplemented with 0.5\% (wt/vol) Trehalose (Sigma Aldrich, T5251) and 0.01\% (wt/vol) Tween (Fisher Scientific, PRH5152). A 150-$\mu$L protein solution is loaded into the custom bioprinter reservoir, ensuring contact with the printer’s ultrasound transducer. To achieve precise control over protein quantity, $\sim$ 2-pL droplets are ejected onto epoxy-silane-coated nanoantennas, forming an array. The samples are dry-incubated at room temperature for 1 hour to enable covalent immobilization of capture antigens in a microarray format. After incubation, three washes with diluted PBST 1x (Thermo Fisher, 28352) are performed, each lasting 5 minutes, to remove any physisorbed or loosely bound proteins. The samples are subsequently rinsed with milli-Q water to prevent salt crystallization and dried using argon gas. Finally, the samples are incubated in a 1 mM m-PEG-amine solution (Biopharma PEG, MF001005-350) prepared in PBS 1x for 15 minutes to backfill unbound sites on the GLYMO layer, thus minimizing nonspecific binding. After brief rinses with PBS 1x and milli-Q water, the samples are dried with argon gas and stored at 4 $^\circ$C for short-term use. For binding kinetics measurements, the functionalized optical chip is integrated into a microfluidic cell (Microfluidic ChipShop, see Supplementary Fig. 11, or Grace Bio-Labs) and positioned on a microscope stage. The microfluidic system also includes a syringe pump and tubing for the injection of analytes and buffers. Antibodies of interest, including anti-SARS-CoV-2 antibody (AcroBiosystems, SAD-S35), anti-Influenza A antibody (AcroBiosystems, HA2-Y198), anti-Influenza B antibody (Thermo Fisher, MA5-29901), Cetuximab (Thermo Fisher, 700308), and Herceptin (Thermo Fisher, MA5-42305), are reconstituted and serially diluted to achieve the desired concentrations, similar to the preparation of antigens. PBS 1x is used as the running buffer for all real-time measurements. Spectral acquisitions are collected at 2-second intervals, and measurements are performed under a continuous flow rate of 15 $\mu$L/min. For the epitope binning experiments, we utilize the H5N1 subtype of avian Influenza A HA (AcroBiosystems, HA1-V52H9), along with four anti-HA antibodies: anti-HA Ab \#1 (Sigma-Aldrich, SAB3501219), anti-HA Ab \#2 (Sigma-Aldrich, SAB3501220), anti-HA Ab \#3 (Sigma-Aldrich, SAB3501221), and anti-HA Ab \#4 (Sigma-Aldrich, SAB3501224). These proteins are reconstructed following the established protocol. Additionally, glycoengineered Cetuximabs, produced through independent manipulation of the Fab and Fc N-glycans via a chemoenzymatic method \cite{wang2013general}, are included in the study. All glycoengineered variants are characterized using liquid chromatography-mass spectrometry to confirm glycan transfer (see Supplementary Fig. 14).

\subsection*{Quantification of deposited protein molecules} 

There exists a one-to-one correlation between the mass (and number) of protein molecules deposited onto a nanoantenna droplet and the concentration of the solution loaded into the bioprinter. For example, in a single droplet ($\sim$ 2 pL) of a 26.5 $\mu$g/mL receptor-binding domain (RBD) solution (with a molecular weight of 26.5 kDa), the resulting deposited mass would be 53 femtograms, corresponding to approximately 1.2$\times$10$^6$ molecules. Measurements obtained from fluorescence microscopy indicate that a single deposited droplet typically occupies an area of approximately 300 $\mu$m$^2$ of the sensor surface. Assuming a uniform distribution of molecules within the droplet, the mass of protein deposited per square micrometer is approximately 176 attograms, equating to around 4,000 molecules per square micrometer. To achieve functionalization of a nanoantenna with a concentration of 1 nM RBD, this requires the deposition of two droplets. It is important to note that the density of deposited proteins in each experiment is influenced by various environmental parameters, including evaporation rate, temperature, and humidity, as well as experimental factors such as the type of protein, protein concentration, viscosity of the solution, sampling variations, and protein aggregation. Also, a Poisson probability distribution governs the randomness associated with the number of ejected molecules in a droplet.

\subsection*{Fluorescence microscopy}
Fluorescence experiments are performed after varied trials of antigens binding using both bulk and acoustic ADE-based functionalization methods. Experiments are performed with an array of biotinylated target antigens tagged with fluorescently-labeled streptavidin, including Alexa Fluor 568 (Invitrogen S11226), Alexa Fluor 647 (Invitrogen S21374), Alexa Fluor 750 (Invitrogen S21384), Alexa Fluor 405 (Invitrogen S32351), and Alexa Fluor 488 (Invitrogen S11223). Dried samples are imaged using the Zeiss AxioImager Fluorescent Microscope and imaged using a Zeiss Axiocam 506 mono camera. Large area images are acquired using the tiling and stitching function available using the Zeiss Zen Pro software package. 

\subsection*{Acknowledgements}
The authors thank Dr. Jack Hu and Dr. Anthony Robert Prudden for insightful discussions and comments. S.A., P.M., and J.D., acknowledge support from the Q-next grant under award no. DE-AC02-76SF00515, which supported the chip development, as well as support from the Chan Zuckerberg Biohub, San Francisco for bioassay development. F.S. acknowledges salary support under NSF grant 1933624.  D.O.L. acknowledges the support of the Knight-Hennessy scholarship for graduate studies at Stanford University. N.A.G. thanks the German National Academy of Sciences Leopoldina for their support via the Leopoldina Postdoctoral Fellowship (LPDS2020-12). S. D. acknowledges support from the Department of Defense (DoD) through the National Defense Science and Engineering Graduate (NDSEG) fellowship and the MURI (grant number N00014-23-1-2567). G.J.B acknowledges support from National Institute of Allergy And Infectious Diseases of the National Institutes of Health under Award Number R01 AI165692. Part of this work was performed at the Stanford Nano Shared Facilities and Stanford Nanofabrication Facilities, which are supported by the National Science Foundation and National Nanotechnology Coordinated Infrastructure under awards ECCS-2026822 and ECCS-1542152.

% \subsection*{Author contributions}
% F.S., S.A., and J.A.D. conceived and designed the study. S.A. and D.O. executed the modeling and numerical simulations with assistance from V.D. S.A. and D.O. carried out the surface functionalization, bioassay, and acoustic bioprinting experiments, with support from F.S., P.M., and K.C. S.A. performed fluorescence microscopy. S.A., D.O., O.H., and S.D., fabricated samples with help from H.C.D., P.M., and V.D. O.H. and S.A. carried out SEM characterization. N.A.G., S.D., and V.D. contributed to the optical measurement of nanophotonic structures. G.B. supplied the glycosylated antibodies. J.A.D. conceived the idea and supervised the project, along with B.T.K., M.L.B., and G.J.B. on the relevant portions of the research. All authors contributed to the preparation and editing of the manuscript.

% \subsection*{Data availability} 
% The data that support the plots within this paper and other findings of this study are available from the corresponding authors on reasonable request. 

% \subsection*{Competing interests} 
% F.S. and J.A.D. are shareholders in Pumpkinseed Technologies, Inc.. The remaining authors declare no competing interests.

\clearpage
\newpage

\bibliography{apssamp}

\end{document}